\documentclass[]{tPHM2e}
 
 \usepackage{amssymb}
  \usepackage{amsmath}
  \usepackage{dsfont}
  \usepackage{color}
  \usepackage[table]{xcolor}
  \usepackage{hyperref}
  \usepackage{graphics}
  \usepackage[]{graphicx}
  \usepackage{rotating}
  \usepackage{array}
  \usepackage{todonotes}
  \usepackage{ulem}
  \usepackage[]{epstopdf}
\newcounter{todocounter}

\begin{document}
\doi{10.1080/14786435.20yy.yyyyyy}
\issn{1478-6443}
\issnp{1478-6435}
\jvol{00} \jnum{00} \jyear{2012} %\jmonth{21 December}

\markboth{Taylor \& Francis and I.T. Consultant}{Philosophical Magazine}

%\articletype{GUIDE}

\title{
Self-Referential Order
 } 

\author{
T. Aste$^{1,2,3}$, 
P. Butler$^{2}$,
and
T. Di Matteo$^{4}$
\vspace{0.5cm} \\
$^1$ Department of Computer Science, UCL Gower Street - London - WC1E 6BT, UK.\\
$^2$ School of Physical Sciences, University of Kent, UK.\\
$^3$ Applied Mathematics, Research School of Physics and Engineering, The Australian National University, Canberra ACT 0200, Australia.
$^4$ Department of Mathematics, King's College London, The Strand, London, WC2R 2LS, UK.
\\ 
%$\ast$ E-mail corresponding author: tomaso.aste@anu.edu.au
}

\maketitle

\begin{abstract}
We introduce the concept of {\it self-referential order} which provides a way to quantify structural organization in non crystalline materials. 
The key idea consists in the observation that, in a disordered system, where there is no ideal, reference, template structure, each sub-portion of the whole structure can be taken as reference for the rest and the system can be described in terms of its parts in a self-referential way. 
Some of the parts  carry larger information about the rest of the structure and they are identified as {\it motifs}.
We discuss how this method can efficiently  reduce the amount of information required to describe a complex disordered structure by encoding it in a set of motifs and {\it matching rules}.
We propose an information-theoretic  approach to define a {\it self-referential-order-parameter} and we show that, by means of  entropic measures, such a parameter can be quantified explicitly.
A proof of concept application to equal disk packing is presented and discussed. 
\bigskip

\begin{keywords}
Self-Referential Order, Disordered structure, information theory, order, structural encoding.
\end{keywords}\bigskip
\end{abstract}

\section{Introduction}
Complex, non-crystalline materials are everywhere and the capability of understanding and mastering disordered atomic packings is crucial to enhance properties of materials. 
The quest for understanding the internal structure of matter has been central to human curiosity since the beginning of science and, despite the remarkable achievements obtained since the  Platonic theory of matter (Timaeus $\sim 360$ BC), still we are only able to describe the structure of a very special class of materials where regular periodic (or quasi-periodic) arrangements of atoms are present. 
However, disorder is not randomness and nor it is a defective, degenerate form of order, real disordered structures show high degrees of organization that can propagate hierarchically through the material. 
Nonetheless, these structures do not present any periodic, predictable pattern and the absence of such regularity is precisely what makes disorder difficult to describe and encode in a way that is both accurate and compact. 
Science is measurement, but disorder is difficult to quantify. 
For instance, in an ordered, crystalline, system one can introduce a quantity called {\it `order parameter'} that measures how close the system is to the perfect crystalline reference structure. 
This parameter is extremely useful to predict the properties of the material.
But in a disordered system, there isn't a unique, ideal, reference structure and a simple parameter that quantifies the kind and amount of disorder cannot be used.
Even  common language lacks of terms to define non-ordered structures. 
Indeed, we are limited to the use of negative identifications: disorder (i.e. disturbance of order) or amorphous (i.e. absence of shape). 
Such a lack of vocabulary is probably  consequence of the fact that the classification of crystalline structures has been one of the great success stories of modern science which has induced us to overlook the evidence that ``disordered'' and ``amorphous'' materials are everywhere and the mastering of good techniques to describe atomic disorder is crucial to enhance material performances. 
It has been recently shown that `order' in amorphous structures can be identified by looking at `patches' that repeat more often than typical \cite{kurchan2011order}.
This approach reveals diverging correlation lengths at glass transition \cite{PhysRevLett.107.045501} shading light on the relations between thermal glass transition and athermal jamming of discrete matter \cite{biroli2013perspective}.
In this paper we follow a similar approach to \cite{kurchan2011order} using  information-theoretic methods to quantify, in a self-referential way, an ordered parameter and identifying the locally most referential structures. 

There are two main technical challenges that have so far slowed down the progress in this field. 
The first has been the lack of experimental data. 
Indeed, until recently, diffraction techniques have been the main experimental tools to study atomic structures inside the bulk of materials. 
However, diffraction gives insights only on the average relative positions of the constituents and the reconstruction of the structure from diffraction data becomes very hard in absence of regularly repeated local units. 
Now, for the first time, atomic-scale tomography techniques are providing us a way to directly ``see'' the complex atomic architectures inside materials. 
Indeed, in the last few years, techniques such as Atom Probe Tomography and Electron Tomography have started to provide direct information about the position of millions of atoms in the bulk of materials \cite{ronhovde2012detection,van2011three,marceau2011quantitative,gault2010spatial,miller2009atom,miller2005atom}. 
In the next few years we will witness a large production of experimental data concerning large-scale complex atomic aggregates.
However, this brings up the second technical challenge concerning the huge size of data to process demanding the development of specific tools and a novel theoretical framework for their interpretation and use.  
Indeed, in  absence of a compact way to encode structural complexity, the processing of this amount of information is still beyond the capability of the world's largest supercomputers.
The total world information storage capacity, currently estimated $10^{20}$ bits, would not be enough to encode the structure of a gram of disordered matter.
%Indeed, despite the fact that  the capability of handling and processing very large datasets has reached remarkable benchmarks (,) and it is growing continuously, the direct processing of the full information concerning a macroscopic piece of matter is still computationally unfeasible. 
There is therefore a demand to develop a general approach to encode complex structures and reduce the amount of information to the relevant part related to the material's functional properties.
In principle, in a disordered material  positions, properties and the interactions of every atom must be recorded independently. 
In some special cases, when the structure is a regular periodic repetition of identical parts (i.e. crystals), the problem can be reduced to the study of the unit cell: a local sub-structure that repeats periodically in space, however this cannot be directly extended to non-crystalline materials. 
Nonetheless, even in these `disordered' materials, geometrical, physical and chemical laws impose local regularities that spontaneously develop into a structural organization spanning the whole system. 
In this paper we show that these regularities can be identified as a set of local motifs that combine together into a hierarchically organized space-filling complex network in a analogous way as an alphabet combines into words which assemble into phrases forming the whole text. 
Retrieving the `alphabet', identifying the `words', uncovering the `grammatical' rules and, ultimately, decoding the `syntax' is the key to describe the structure of non-crystalline matter.

\section{Describing the structure in terms of itself: self-referential order}
The key-idea at the basis of the present work is very simple: in the absence of a pre-definite template reference structure, we can use a part of the material as a reference structure for another part. 
The structure is consequently encoded with a {\it self-referential} description.
Indeed, from a general information-theoretic perspective, the identification of the unit cell of a crystalline structure is a very efficient way to encode a structure with the amount of data required to encode the structure passing from $O(n)$ to $O(1)$.
Even in the absence of any previous knowledge of crystallography it is still rather straightforward to identify the unit cell from the information about the positions of all atoms.  
Indeed, it is sufficient to take a portion of the structure, translate it in space and see when and where it perfectly overlaps with another part of the structure. 
The smallest portion of the structure that periodically overlaps with all other parts of the structure is the unit cell.
In the context of this paper this is the simplest case of self-referential description where only one local motif -the unit cell- is sufficient to entirely describe the whole crystal.

  \begin{figure}[t]
  \includegraphics[width=0.95\columnwidth]{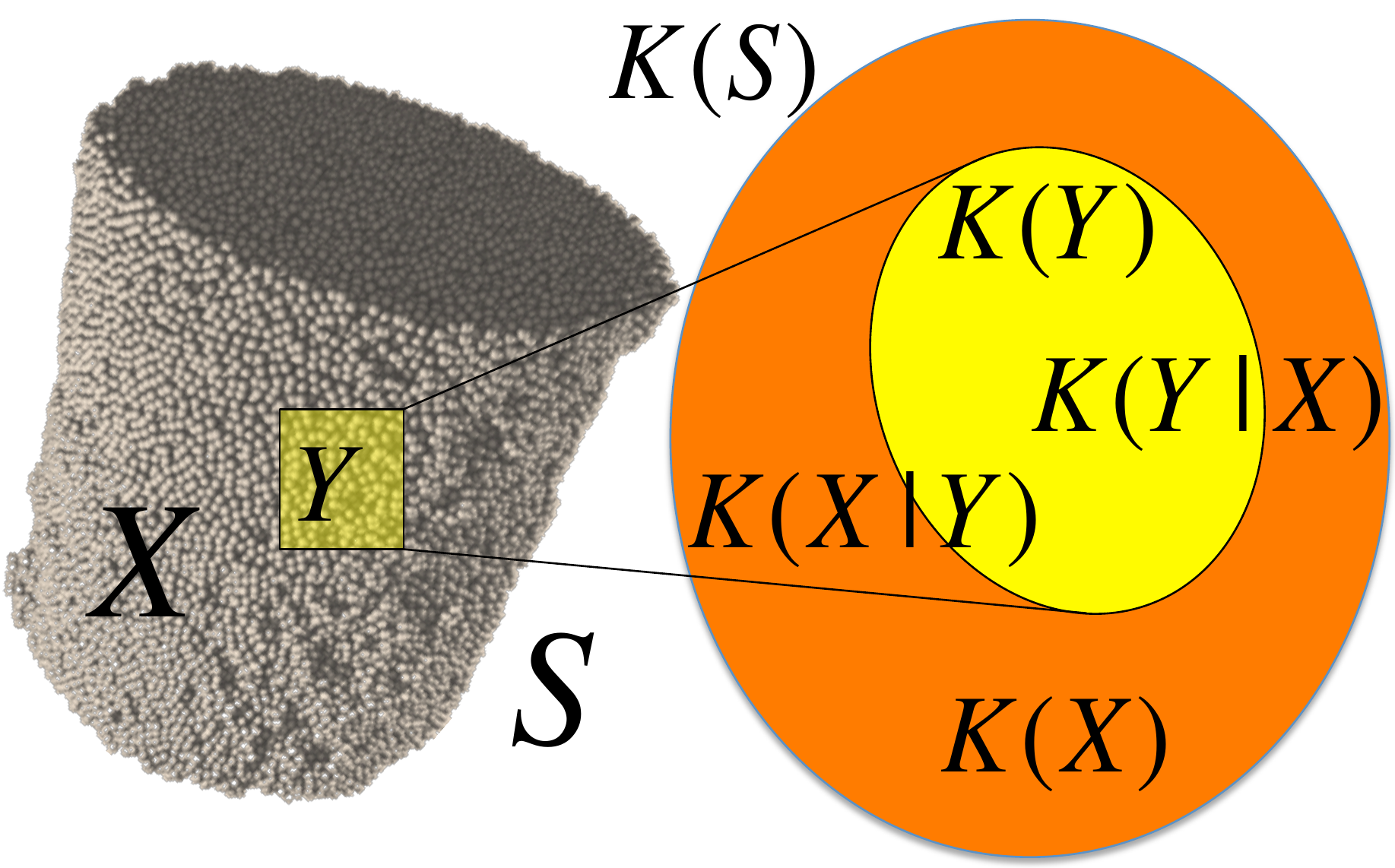}
        \caption%[PMFG]{\scriptsize 
        {\label{f1}
In absence of a pre-defined template reference structure, one can use a portion ($Y$) of the structure to describe the whole structure $S=X\cup Y$.
The knowledge about the portion $Y$  can reduce the uncertainty about the rest of the structure $X$.
The Kolmogorov complexity, here denoted with $K(X)$ and $K(Y)$, measure the information contained in $X$ and $Y$ respectivelly.
For instance, in the case in which the rest of the structure $X$ is completely determined by the knowledge of the portion $Y$, we have $K(S) = K(Y)$.
In this case, the conditional information about $X$ give $Y$,  $K(X | Y) $, is equal to zero.
        }
  \end{figure}

\subsection{An ideal approach}
Let us consider a structure $S$ and let us consider it as composed of a large portion $X$ and a smaller portion $Y$, so that $S=X\cup Y$.
To measure the amount of self-referential order we need to be able to quantify how the knowledge about the portion $Y$ reduces the amount of information needed to encode $X$.
Formally we need a measure of information content such as the Kolmogorov complexity $K$ \cite{Kolmogorov1968,Solomonoff1964,Solomonoff1960,Li1993}.
In simple terms, the quantity $K(X)$  is the amount of information necessary to  describe  $X$.
Its conditional counterpart, $K(X|Y)$, is the amount of information necessary to describe  $X$,  given the full knowledge of  $Y$.
When the knowledge about a portion $Y$ of the structure is sufficient to describe the rest of the structure we must have $K(S)=K(Y)$ and $K(X|Y)=0$.
Conversely, when the knowledge about a portion does not add any knowledge about the rest of the structure we must have $K(X|Y)=K(X) > 0$.%\footnote{ $X \setminus Y$ is the set difference between $X$ and $Y$, i.e. $ X \setminus Y = \{r \in X | r \notin Y\}$.}
%Its conditional counterpart: $K(X|Y)$ is the amount of information necessary to describe structure $X$,  given the full knowledge its part  `$Y$'.
%% portion $Y$ and compare it to other portions or to the whole structure (i.e. Kolmogorov complexity) $K(Y)$ which is the amount of information necessary to describe structure `$Y$'.
%%We then need also its conditional counterpart: $K(Y|Y)$ which is the amount of information necessary to describe structure `$Y$' given the full knowledge of structure $Y$.
%We have:
%\begin{equation}
%K(X | Y)= \begin{cases} 
%0&\mboy{if $X$ is completely determined by $Y$} \\
%K(X / Y) & \mboy{if $Y$ provides no useful information }. 
%\end{cases}		
%\end{equation}

We could therefore introduce the {\it self-referential order parameter}
\begin{equation}
s\!_{_X}\!(Y)= 1- \left( \frac{K(X | Y)}{K(X)}  \right) \;\;,
\label{m_K}
\end{equation}
which is equal to one if the system is fully self-referentially ordered and it is equal to zero if completely random. 
This approach formally defines {\it self-referential order} and it would  solve the problem.
However, --unfortunately-- Kolmogorov complexity is not a computable quantity.
 
 \subsection{The entropic way}
A quantity that measures information content is the {\it entropy} that, in the Shannon formulation \cite{Shannon1948}, can be  written as:
\begin{equation}
H(X) = - \sum_{r\!_{_X}} p\!_{_X}(r\!_{_X}) \log_2 p\!_{_X}(r\!_{_X})\;\;,
\label{H}
\end{equation}
where $p\!_{_X}(r\!_{_X})$ is the probability of occurrence, in $X$, of a configuration with a given set of structural properties, denoted with $r\!_{_X}$.
Entropy is everywhere in physics; it is a thermodynamic state variable and the Shannon formula coincides with the Gibbs derivation (with base-$e$ log and multilplied by $k_B$ \cite{chandler1987introduction}) of the entropy for the canonical ensemble.  
Here we shall use entropy for its information significance: $H(X)$ is the amount of information encoded into a structure $X$ when its properties $r\!_{_X}$ are considered.
We then use entropic measure of information instead of the Kolmogorov complexity.
In analogy with the previous section we can therefore look for the information about $X$ provided by the knowledge of $Y$.
%When the portion $Y$ encodes the full information about $X$ then $H(X)=H(Y)$.
The remaining entropy of variable $X$ when variable $Y$ is known is quantified by the conditional entropy $H(X|Y)$. 
Therefore, an entropic measure of self-referential order is:
\begin{equation}
s\!_{_X}\!(Y) = 1- \frac{ H(X|Y) }{ H(X)} \;\;.
\label{m_H}
\end{equation}
We have $0\le H(X|Y)\le H(X)$, therefore this quantity is defined in the interval $0 \le s\!_{_X}\!(Y) \le 1$ where 0 is associated to a  random state and 1 is instead observed for perfect self-referential order.
We can use the identity $H(X|Y) = H(X,Y)-H(Y)$ obtaining the equivalent expression
\begin{equation}
s\!_{_X}\!(Y) = 1- \frac{ H(X,Y)-H(Y) }{ H(X)} \;\;,
\label{m_Hbis}
\end{equation}
which also reads
\begin{equation}
s\!_{_X}\!(Y) =  \frac{ H(X)+H(Y) -H(X,Y)}{ H(X)} \;\;.
\label{m_Hbis1}
\end{equation}
One may notice that the quantity on the numerator is the mutual information: $I(X;Y)=H(X)+H(Y) -H(X,Y)$,  therefore this measure quantifies the relative mutual dependence between  structures $X$ and $Y$.

\section{Motifs}
There must be parts of the structure that carry larger amount of information about the whole structure with respect to others. 
These high information-content portions are repeated similarly in the structure more often that others and therefore they are of particular relevance.
We look for local sub-structures containing maximal relative information. 
We shall call them `motifs' these are equivalent to the `patches' used in \cite{kurchan2011order}.
In general, more than one motif is necessary to encode a disordered structure.
Furthermore, these motifs do not repeat perfectly across the structure and therefore they must be described in statistical terms. 
Motifs are the set of local structures from which the whole structure can be most efficiently encoded.
Frequency of occurrence, fluctuations and  relations between  motifs characterize and quantify the kind and amount of disorder in the structure.
%The methodology to identify these motifs is based on an information-theoretic approach where we quantify the amount of information that a given sub-part of the structure provides to describe the remaining part of the structure. 
We then use these motifs as an encoding alphabet and we search for an efficient description of the entire structure with the shortest code-length.
By identifying the recurrent structural motifs and by uncovering the rules governing their combination into a space-filling network, we can encode the structure of complex materials into a compressed format.

%\subsection{Find structural motifs}
Motifs can be identified from Eqs.\ref{m_K} or \ref{m_H} by looking at the local parts that maximally contribute to the information about the whole structure, i.e. the portions $Y$ associated with the largest $s\!_{_X}\!(Y)$.
%We  developed an unbiased feature-detection procedure to find local motifs based on conditional entropy. 
%Entropy is the statistical mechanics instrument to quantify disorder and it is used in information theory to measure the amount of information. 
%Conditional entropy quantifies the remaining entropy (i.e. remaining uncertainty) of a variable when the value of another variable is known. 
%For example, if a portion, $M$, of the material contains the full information to encode another portion, $P$, then their conditional entropy $H(P |M)$ must be zero; conversely if the structural information in $M$  is of no use to encode $P$  then $H(P | M)$ is maximal and equal to $H(P)$, which is the entropy of  $P$.
%We can take a local portion of the material, allow for some tolerance for fluctuations, and then measure the conditional entropy, $H(P |M)$, between such a local motif $M$ and other parts of the system $P$. 
%The difference
%\begin{equation}
%I(P,M) = H(P)- H(P |M)						
%\end{equation}
%quantifies the reduction in uncertainty about $P$ produced by the knowledge of $M$. 
%This quantity is called {\it mutual information}. 
%We can identify the structural motifs by searching for small local motifs $M$ with minimum entropy $H(M)$ but with mayimum mutual information $I(P ,M)$. 
%If, for instance, we apply this reasoning to a crystal, we will soon discover, without any assumption or prior knowledge, that a small part of the material -the unit cell- contains the whole information about the system. 
%\subsection{Measure their statistical recurrences}
Once the motifs are identified, one must quantify their recurrence in the structure.
This can be done in three steps: 
(i) count the relative frequency of occurrence of each local motif; 
(ii) compute the probability distribution of its fluctuations; 
(iii) estimate the entropy. 
A computationally fast identification of the motifs in  presence of structural fluctuations is a very challenging task. 
Another challenge is associated with possible overlaps between motifs that make their unique identification ambiguous and requires the introduction of ``exclusion rules'' (i.e. when two motifs overlap, only one must be counted at the time) and statistical enable analysis (i.e. all encodings resulting from the different exclusions) must be considered.

%\subsection{Matching rules} 
Motifs are building blocks that connect to each-other forming a space-filling three-dimensional structure. 
When described in terms of motifs, the structure is characterized by two aspects: 
(1) topology - a network of interconnected motifs; 
(2) geometry, where position and orientation of each motif is specified. 
Due to the possible overlaps between motifs, there can be more than one network for a given structure, the ensamble all these networks must be considered. 
For a given network, the matching rules can be identified from a statistical study of local co-occurrences. 
Matching rules are both topological and geometrical. 
Indeed, motifs can join together only in specific relative positions and orientations.

The description of a structure in terms of the network of motifs and their matching rules provides a compact encoding of the structure. 
For example, a crystal is reduced to only one motif (a parallelepipedal unit cell), one  topological matching rule (6 neighbors) and one  geometrical matching rule (unit cells join by opposite faces). 
In general, for a complex structure we have a large -but finite and non-scaling- number of motifs $m$ and a order $O(m^2)$ of matching rules. 
Therefore the amount of information required to encode the structure is of the order  $O(m^2)$. 
A-priori it is quite hard guesswork to estimate the size of $m$, which -of course- varies from system to system. 
The experience acquired with disordered sphere packings \cite{delaney2010combining,aste2008structural,AsteLoacalGlobal06} suggests us that in these systems $m$ is of the order of $10^2$, and the matching rules are of the order of $10^4$ (note that resolving all the reciprocal orientations can be demanding). 
This may seem a large number but it must be pointed out that in terms of information compression we are passing from an information size of 
the order of $10^{20}$ (hundreds of billions of gigabytes), which is certainly beyond computable sizes, to a size of $10^4$ bytes (tens of kilobytes), which is computationally insignificant.
Furthermore, for many practical purposes, a precise definition of the local geometrical configuration and its orientation is often irrelevant and the information can therefore be further reduced. 
Let us here explain this point with an example from the results on sphere packings in \cite{Anikeenko08} where it was shown that local tetrahedral motifs are related to the description of a structural transition at the Random Close Packing limit. 
In this case, one can identify $m=2^6=64$ motifs (corresponding to the number of different tetrahedra that can be build with long/short edges) which correspond to $(4\times2^6)^2/2\sim33,000$ matching rules (the probabilities to merge together the different tetrahedra face by face). 
However, it can be shown that two motifs (open/closed tetrahedra) and one matching rule (probability to join together two closed tetrahedra) are sufficient to fully describe such a transition.  
%spot from bulk observations.
%From a general perspective we can view a crystal as a very efficient way to encode the massive information of a three dimensional bulk structure into only a few parameters associated to the unit cell.
%The unit cell is an elementary motif that can be simply translated periodically into space to generate any structure of any size.
  \begin{figure}[t]
  \center
      \includegraphics[width=1.0\columnwidth]{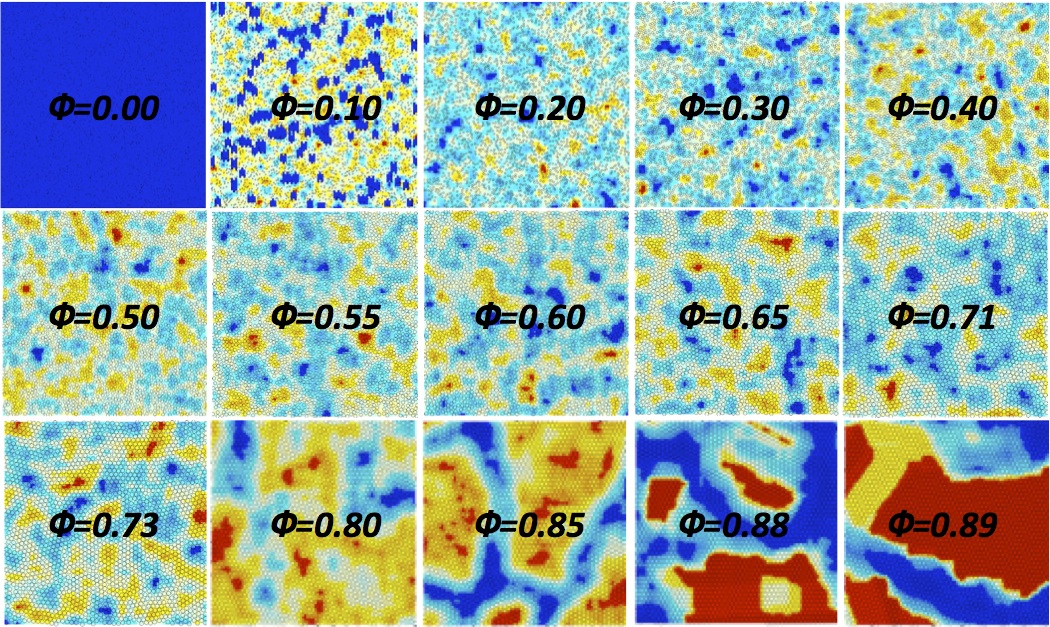}
        \caption%[PMFG]{\scriptsize 
        {\label{fSnapShots}
   Snapshots of the local self-referential order parameter $s_X(Y)$.
   The local portion $Y$ is a square of edge 5 disc diameters.
 The pictures are a heat map (blue low red high, color online) representing the relative values of $s_X(Y)$ for a portion centered in each given part of the packing.
 $\Phi$ indicates the packing fraction of each sample.
 Colormap is rescaled for each image.
        }
  \end{figure}
  
   \begin{figure}[t]
  \center
      \includegraphics[width=0.75\columnwidth]{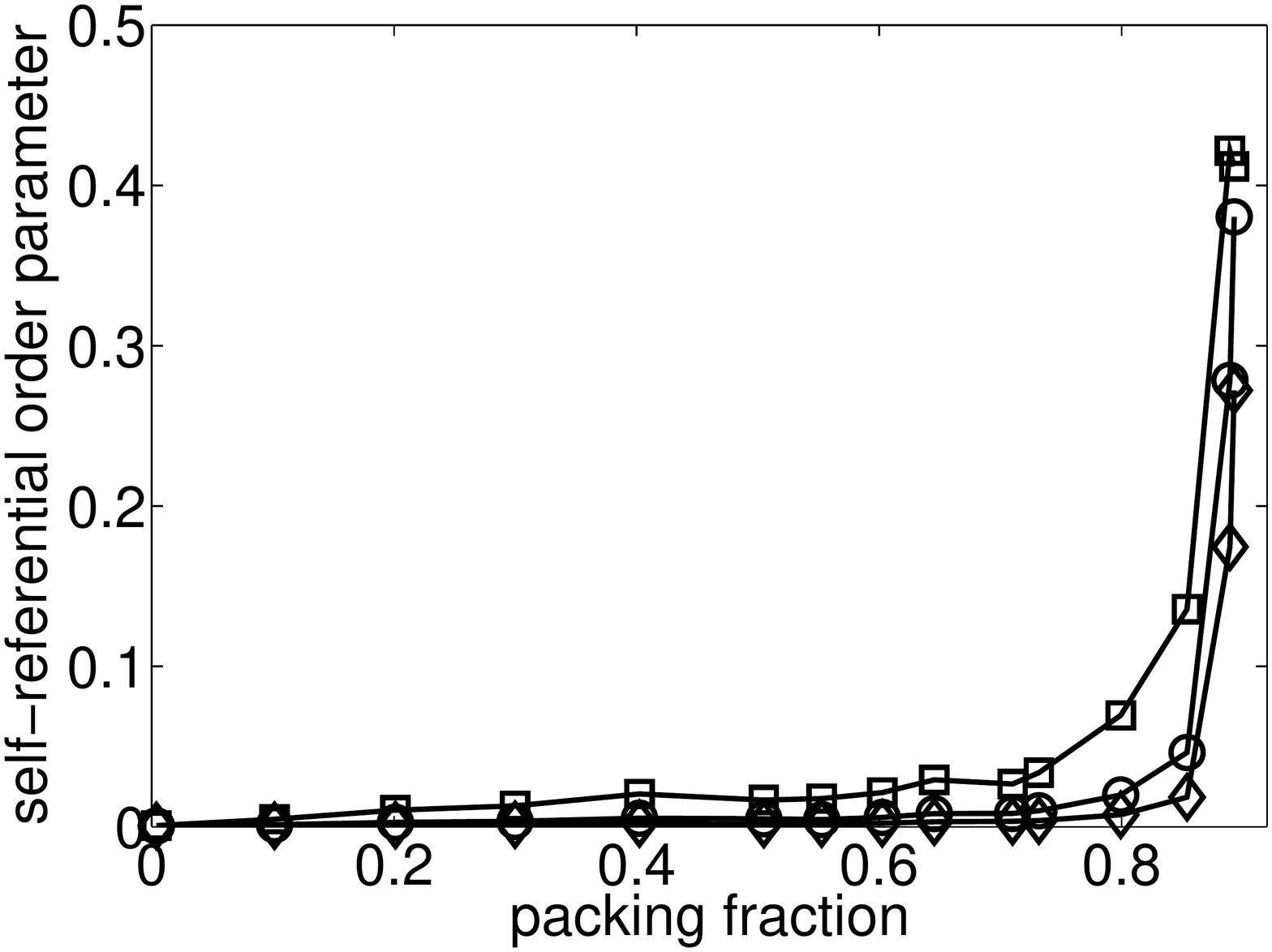}\\
 \includegraphics[width=0.75\columnwidth]{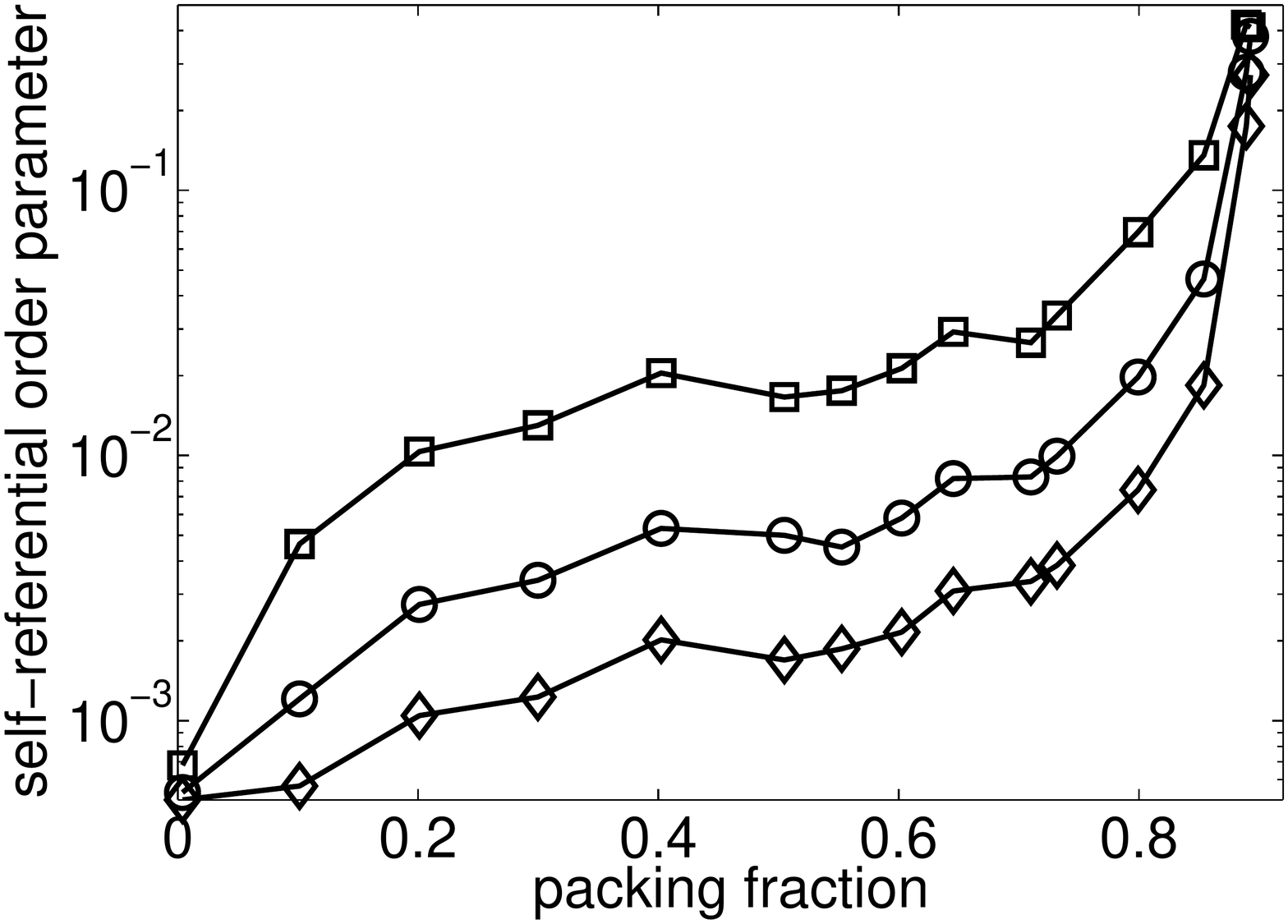}
       \caption%[PMFG]{\scriptsize 
        {\label{fGlobal}
  Global values of the self referential order parameter $\hat s$ vs. packing fraction displayed in both linear and semi-logarithmic scale.
  Different curves ($\diamond$, $\circ$ or $\square$ symbols) correspond to different sizes of the local portion $Y$, which are squares respectively with edges equal to 3, 5 or 10 disk-diameters. 
        }
  \end{figure}

  \begin{figure}[t]
  \center
      \includegraphics[width=0.85\columnwidth]{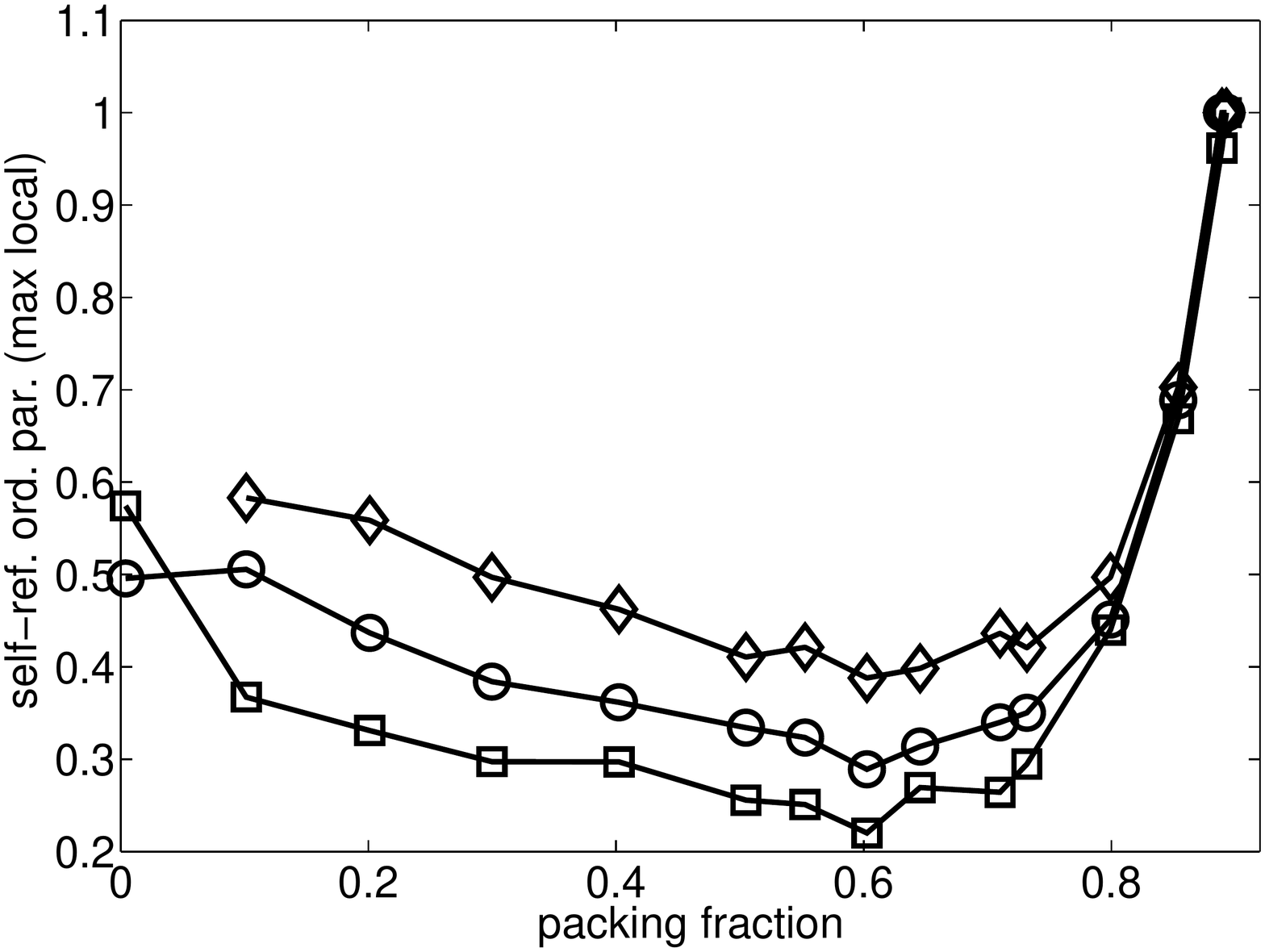}
        \caption%[PMFG]{\scriptsize 
        {\label{fLocal}
        Average maximum local values of the self referential order parameter for each sample.
        The average is over the 10\% largest $s_X(Y)$.
  Different curves ($\diamond$, $\circ$ or $\square$ symbols) correspond to different sizes of the local portion $Y$, which are squares respectively with edges equal to 3, 5 or 10 disk-diameters. 
         }
  \end{figure}

 %  \begin{figure}[t]
%  \center
%      \includegraphics[width=0.85\columnwidth]{FigDiversity.pdf}
%        \caption%[PMFG]{\scriptsize 
%        {\label{fDiversity}
%  Fraction of local 
%        }
%  \end{figure}

 \section{Results}
In this paper we report a preliminary investigation about the quantification of self-referential order in two-dimensional disks packings generated via molecular dynamic  simulations.
The results presented here are a `proof of concept' demonstrating that this method can be used quantitatively. 
Extended applications to three dimensional structures from simulations and experiments are under investigation.

We generate several packings of disks at various packing fractions by using the algorithm proposed by  \cite{Skoge06}, which is a molecular dynamic simulation with constant compression rate. 
We terminate the simulation when a desired packing fraction is reached, before the reach of (local) jamming.
We report results for 15 samples comprising 5,000 disks representing a range of packing fractions between 0 to $\sim 0.9$.

We compute the self referential order parameter by looking at the Vorono\"{\i} volumes  around each disk and identifying a set of $m=500$ kinds of motifs classified in terms of their different volumes.
We verify that the method is robust against this choice with analogous results obtained for $m=100$ or $m=2,000$.
We then take a local square portion $Y$ of the sample and compute $s_X(Y)$  by applying Eq.\ref{m_Hbis1}.
We repeat the process in 10,000 different portions regularly displaced across each sample.  

In Figure \ref{fSnapShots}  distributions of the local self-referential order parameter $s_X(Y)$ inside each sample and across the samples are shown.
One can note that the values are low at low packing fractions where the system is essentially in a random state. 
Conversely, they are large at high packing factions where the system  starts nucleating crystalline regions.
This is quantified and shown in  Fig.\ref{fGlobal} where we report a global measure of self referential order parameter ($\hat s_X(Y)$) computed by estimating the joint probability to have given fractions of Vorono\"{\i} volumes simultaneously present in any of the portions $Y$ and in the rest of the sample $X=S\cap Y$. 
One can see that the self referential order parameter increases with packing fraction to reach a maximum at the largest packing of $\Phi\sim 0.9$.
From the semi-log plot in Fig.\ref{fGlobal} we can note that this parameter ranges over 4 order of magnitude, with an interesting plateau appearing between packing fractions $\sim 0.4$ and $\sim 0.7$. 
Let us note that the largest packing fraction attainable for equal disks is $\Phi = \pi/\sqrt{12}\simeq 0.907$ \cite{ppp}, which corresponds to a perfectly ordered, crystalline, triangular packing. 
Our densest packing has still some defects that lower slightly its packing fraction.
These defects are clearly visible in  Fig. \ref{fSnapShots} where one can appreciate that in correspondence with miss-alignment of the crystalline order we observe lower values of $s_X(Y)$.
Indeed, these defective regions are less representative of the sample.
We can also note that, conversely, at lower packing fractions the most representative local portions are not compact configurations with crystalline symmetry but rather more complex and less compact configurations.
In general, at different packing fractions different local configurations carry more or less information about the rest of the sample structure.  
We investigated the presence of highly referential motives by looking at the maximum values of $s_X(Y)$ in each sample.
Specifically, we quantified the portions of sub structures carrying the largest information by identifying the 10\% largest  $s_X(Y)$ per each sample.
In Fig.\ref{fLocal}  we show the values of the average self referential order parameter $s_X(Y)$ in this top 10\% subset of most representative configurations.
One can note that at large packing fractions, where the structure is essentially crystalline, only few configurations carry all structural information. 
Interestingly, also at very low packing fractions, where the structure is essentially random, again a small part of the most informative configurations characterize well the whole structure.  
On the other hand, at intermediate packing fractions -around $\Phi\simeq0.6$- the structure is more complex and even the most informative local configurations carry, in average, a smaller amount of information about the rest of the system.

\section{Conclusion}
We addressed the intriguing question concerning how atoms organize themselves inside non-crystalline, complex materials and how to extract, filter and encode this information in an efficient and meaningful way.
To this purpose we introduced the concept of self-referential-order and we proposed a method to quantify it from  entropic measures.
There are over one billion trillion atoms in a gram of matter, and in the absence of a regular, ordered arrangement, the characterization of an amorphous structure would require accounting for the position of every atom. 
This is an impossible task that would require over a billion terabytes.
However, the material functional properties are associated with a much smaller amount of information. 
In this paper we have illustrated a general approach to encode complex structures and to reduce this overwhelming amount of information to the relevant part related to the material's functional properties. 
Our method can be used to select the most informative portions of the material, the `motifs', and encode the complex structure in a set of motifs and matching rules reducing dramatically the amount of information required. 
In this paper we present a `proof of concept' with application to equal disk packing at different packing fractions.
We found that the self-referential-order parameter well characterizes globally the transition towards crystallization, but also it identifies locally the emergence of an increasing complexity at intermediate packing fractions. 
Future studies will be dedicated to the analysis of three dimensional structures from experiments and large scale simulations.  
Our information filtering and encoding techniques can be directly applied to very different kinds of complex structures which are defined in high dimensional phase-spaces: the study of the structure of dependency in financial systems \cite{NJP10,pozzi2013spread} or the structure of gene co-expressions in biological systems \cite{HierInfoFilt2012}.

\bibliographystyle{unsrt}
%\bibliography{/Users/tomaso/TOM/Biblio/GeneralBiblio,/Users/tomaso/TOM/Biblio/SelfReferentialOrder}

\begin{thebibliography}{10}

\bibitem{ronhovde2012detection}
P~Ronhovde, S~Chakrabarty, D~Hu, M~Sahu, KK~Sahu, KF~Kelton, NA~Mauro, and
  Z~Nussinov.
\newblock Detection of hidden structures for arbitrary scales in complex
  physical systems.
\newblock {\em Scientific reports}, 2, 2012.

\bibitem{van2011three}
Sandra Van~Aert, Kees~J Batenburg, Marta~D Rossell, Rolf Erni, and Gustaaf
  Van~Tendeloo.
\newblock Three-dimensional atomic imaging of crystalline nanoparticles.
\newblock {\em Nature}, 470(7334):374--377, 2011.

\bibitem{marceau2011quantitative}
RKW Marceau, LT~Stephenson, CR~Hutchinson, and SP~Ringer.
\newblock Quantitative atom probe analysis of nanostructure containing clusters
  and precipitates with multiple length scales.
\newblock {\em Ultramicroscopy}, 111(6):738--742, 2011.

\bibitem{gault2010spatial}
Baptiste Gault, Michael~P Moody, Frederic De~Geuser, Alex La~Fontaine, Leigh~T
  Stephenson, Daniel Haley, Simon~P Ringer, et~al.
\newblock Spatial resolution in atom probe tomography.
\newblock {\em Microscopy and Microanalysis}, 16(1):99, 2010.

\bibitem{miller2009atom}
Michael~K Miller and RG~Forbes.
\newblock Atom probe tomography.
\newblock {\em Materials Characterization}, 60(6):461--469, 2009.

\bibitem{miller2005atom}
M.K. Miller.
\newblock Atom probe tomography.
\newblock {\em Handbook of Microscopy for Nanotechnology}, pages 227--246,
  2005.

\bibitem{kurchan2011order}
Jorge Kurchan and Dov Levine.
\newblock Order in glassy systems.
\newblock {\em Journal of Physics A: Mathematical and Theoretical},
  44(3):035001, 2011.

\bibitem{PhysRevLett.107.045501}
Fran\ifmmode \mbox{\c{c}}\else~\c{c}\fi{}ois Sausset and Dov Levine.
\newblock Characterizing order in amorphous systems.
\newblock {\em Phys. Rev. Lett.}, 107:045501, Jul 2011.

\bibitem{biroli2013perspective}
Giulio Biroli and Juan~P Garrahan.
\newblock Perspective: The glass transition.
\newblock {\em arXiv preprint arXiv:1303.3542}, 2013.

\bibitem{Kolmogorov1968}
Andrei~Nikolaevich Kolmogorov.
\newblock Three approaches to the quantitative definition of information.
\newblock {\em International Journal of Computer Mathematics}, 2(1-4):157--168,
  1968.

\bibitem{Solomonoff1964}
Ray~J Solomonoff.
\newblock A formal theory of inductive inference. part i.
\newblock {\em Information and control}, 7(1):1--22, 1964.

\bibitem{Solomonoff1960}
Ray~J Solomonoff.
\newblock {\em A preliminary report on a general theory of inductive
  inference}.
\newblock Citeseer, 1960.

\bibitem{Li1993}
Ming Li and Paul~MB Vitanyi.
\newblock An introduction to kolmogorov complexity and its applications.
\newblock 1993.

\bibitem{Shannon1948}
Claude~Elwood Shannon and Warren Weaver.
\newblock A mathematical theory of communication, 1948.

\bibitem{chandler1987introduction}
David Chandler.
\newblock Introduction to modern statistical mechanics.
\newblock {\em Introduction to Modern Statistical Mechanics, by David Chandler,
  pp. 288. Foreword by David Chandler. Oxford University Press, Sep 1987.
  ISBN-10: 0195042778. ISBN-13: 9780195042771}, 1, 1987.

\bibitem{delaney2010combining}
Gary~W Delaney, T~Di~Matteo, and Tomaso Aste.
\newblock Combining tomographic imaging and dem simulations to investigate the
  structure of experimental sphere packings.
\newblock {\em Soft Matter}, 6(13):2992--3006, 2010.

\bibitem{aste2008structural}
T~Aste and T~Di~Matteo.
\newblock Structural transitions in granular packs: statistical mechanics and
  statistical geometry investigations.
\newblock {\em The European Physical Journal B}, 64(3-4):511--517, 2008.

\bibitem{AsteLoacalGlobal06}
T.~Aste, M.~Saadatfar, and T.~J. Senden.
\newblock Local and global relations between the number of contacts and density
  in monodisperse sphere packs.
\newblock {\em J. Stat. Mech.}, P07010, 2006.

\bibitem{Anikeenko08}
A.~V. Anikeenko, N.~N. Medvedev, and T.~Aste.
\newblock Structural and entropic insights into the nature of the
  random-close-packing limit.
\newblock {\em Phys. Rev. E}, 77:031101, 2008.

\bibitem{Skoge06}
Monica Skoge, Aleksandar Donev Frank~H. Stillinger, and Salvatore Torquato.
\newblock Packing hyperspheres in high-dimensional euclidean spaces.
\newblock {\em Phys. Rev. E.}, 74:041127, 2006.

\bibitem{ppp}
T.~Aste and D.~Weaire.
\newblock {\em The Pursuit of Perfect Packing}.
\newblock Institute of Physics, Bristol, 2000.

\bibitem{NJP10}
T~Aste, W~Shaw, and T~Di{~}Matteo.
\newblock Correlation structure and dynamics in volatile markets.
\newblock {\em New Journal of Physics}, 12(8):085009, 2010.

\bibitem{pozzi2013spread}
F~Pozzi, T~Di~Matteo, and T~Aste.
\newblock Spread of risk across financial markets: better to invest in the
  peripheries.
\newblock {\em Scientific Reports}, 3:1665, 2013.

\bibitem{HierInfoFilt2012}
Won-Min Song, T.~Di~Matteo, and Tomaso Aste.
\newblock Hierarchical information clustering by means of topologically
  embedded graphs.
\newblock {\em PLoS ONE}, 7(3):e31929, 2012.

\end{thebibliography}

\end{document}